\documentclass[fleqn,10pt]{SelfArx} 

\usepackage[english]{babel} 

\usepackage{lipsum} 

\definecolor{color1}{RGB}{0,0,90} 
\definecolor{color2}{RGB}{0,20,20} 

\usepackage{hyperref} 

\hypersetup{
	hidelinks,
	colorlinks,
	breaklinks=true,
	urlcolor=color2,
	citecolor=color1,
	linkcolor=color1,
	bookmarksopen=false,
	pdftitle={Title},
	pdfauthor={Author},
}

\JournalInfo{.} 
\Archive{.} 

\PaperTitle{On the use of organic semiconductors as handles for optical tweezers experiments$:$ trapping and manipulating polyaniline (PANI) microparticles} 

\Authors{Kairon M. Oliveira*, Tiago A. Moura, Janaísa L. C. Lucas, Alvaro V. N. C. Teixeira, Márcio S. Rocha, and Joaquim B. S. Mendes**} 
\affiliation{\textit{Departamento de Física, Universidade de Viçosa, 36570-900, Viçosa, MG, Brazil}} 
\affiliation{*\textbf{Corresponding author}: kairon.oliveira@ufv.br} 
\affiliation{**\textbf{Corresponding author}: joaquim.mendes@ufv.br} 

\newcommand{\keywordname}{Keywords} 

\Abstract{Here we propose the use of the organic semiconductor polyaniline (PANI) for the preparation of spherical-shaped microparticles to serve as handles in optical tweezers (OT) experiments. The stable trapping and manipulation of PANI beads was demonstrated for the first time, using a Gaussian (TEM$_{00}$) beam optical tweezers. The trap stiffness was characterized for various different parameters such as the bead radius, the laser power and the distance between the bead and the coverslip of the sample chamber, attesting the viability of using such material for optical manipulation. Since the effective optical properties of PANI can be modulated by the synthesis process, new related applications are also proposed. The results of the present work therefore open the door for using semiconductor polymeric materials in OT applications.}

\begin{document}
\maketitle 
\thispagestyle{empty} 

\section*{Introduction} 

\addcontentsline{toc}{section}{Introduction} 


The use of optical trapping approaches to manipulate micrometer-sized and nanometer-sized objects promoted a revolution in several areas of knowledge since the seminal work of Arthur Ashkin in the late of 1960s \cite{ashkin2001history}. Among some of the main applications we highlight here the trapping of single viruses and bacteria \cite{ashkin1987optical}, single cell manipulation \cite{zhang2008optical}, single molecule force spectroscopy for studying intermolecular interactions \cite{RochaAPLPso, Alves2015, Oliveira2017} and microrheology \cite{gieseler2021optical}. Furthermore, from the last decade optical tweezers (OT) have also become the basis of emerging technologies such as tractor beams \cite{shvedov2014long}, volumetric displays \cite{smalley2018photophoretic}, biosensors \cite{rodriguez2017optical}, micromotors \cite{campos2018topological, campos2019germanium}, and others \cite{polimeno2018optical}. This recent change of paradigm in using optical tweezers is based on two pillars: the generation of more complex trapping configurations employing spatial light modulators that allow the creation of structured beams, such as frozen waves \cite{suarez2020experimental}; and the emergence of applications involving new materials that allow exploring other types of interactions between radiation and matter that are beyond the classical and well studied optical forces (scattering and gradient forces). 

Recently, our group showed that micrometer-sized particles made of topological insulators exhibit a very interesting oscillatory dynamics when manipulated by a traditional Gaussian (TEM$_{00}$) beam optical tweezers. Such dynamics can be modulated by changing simple setup parameters such as the laser power or the distance between the particle and the coverslip used to construct the sample chamber \cite{campos2018topological}. After that work, a similar behavior was also observed for semiconductor particles (germanium \cite{campos2019germanium} and silicon \cite{moura2020silicon}) with an additional feature: the oscillations in this case depend on the polarization direction of the laser beam. This oscillatory dynamics observed for semiconductor and topological insulator beads opened the door for new applications of optical tweezers, \textit{e. g.}, in microrheological studies and in the development of single molecule heat engines \cite{blickle2012realization} and micromotors \cite{quinto2014microscopic}.

\begin{figure*}[ht]\centering 
	\includegraphics[width=\linewidth]{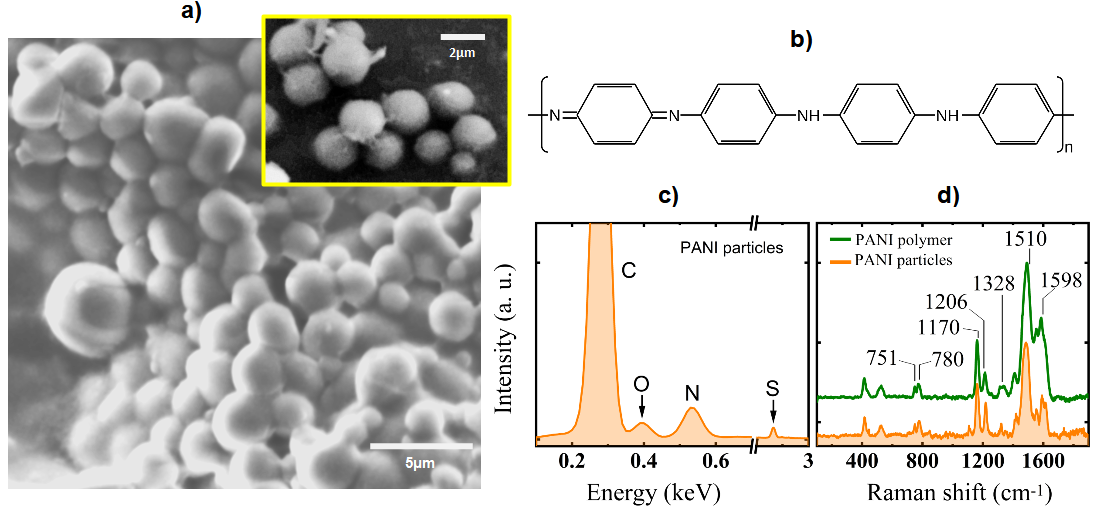}
	\caption{\textbf{(a)} A typical image of the synthesized PANI particles  obtained using scanning electron microscopy (SEM). \textbf{(b)} Molecular structure of PANI at pH = 14 and \textbf{(c)} signal of the atomic components of the PANI particles obtained using Energy-Dispersive Spectroscopy. \textbf{(d)} Raman spectra of PANI particles and PANI polymer in emeraldine base.}
	\label{fig:Fig1}
\end{figure*}

Within this perspective, developing other types of particles, whose optical properties can be modulated in their synthesis processes, becomes fundamental for the construction of new technologies based on optical manipulation. In particular, among the various available materials, the polymer polyaniline (PANI) presents a great alternative by the fact that many of its optical and conductive parameters (charge carrier density, absorption coefficient,  refractive index, reflectance, etc) can be modulated through changes in the pH of the surrounding solution. By changing the pH, in fact, one can induce  changes in the chemical bonds of the polymer structure, modifying its oxidation and thus altering its base \cite{beygisangchin2021preparations, babel2021review}. Furthermore, PANI is a polymer that presents a low cost, easy synthesis, a better stability than other conducting polymers \cite{babel2021review}, and variable properties which allows one to synthesize beads of this material in an appropriate solution, such that these beads can even exhibit a semiconductor character \cite{de2021photoconductivity}. In addition, PANI has interesting electronic and optical properties which makes this polymer very attractive to be used in a large range of applications in electronics, optics, energy storage, environmental sciences, spintronics, and in the protection of metals against corrosion \cite{shi2018synthesis, ding2021electrochemical, wu2021reinforced, mendes2017efficient, mendes2022spin}.

\begin{figure*}[ht]\centering 
	\includegraphics[width=0.9\linewidth]{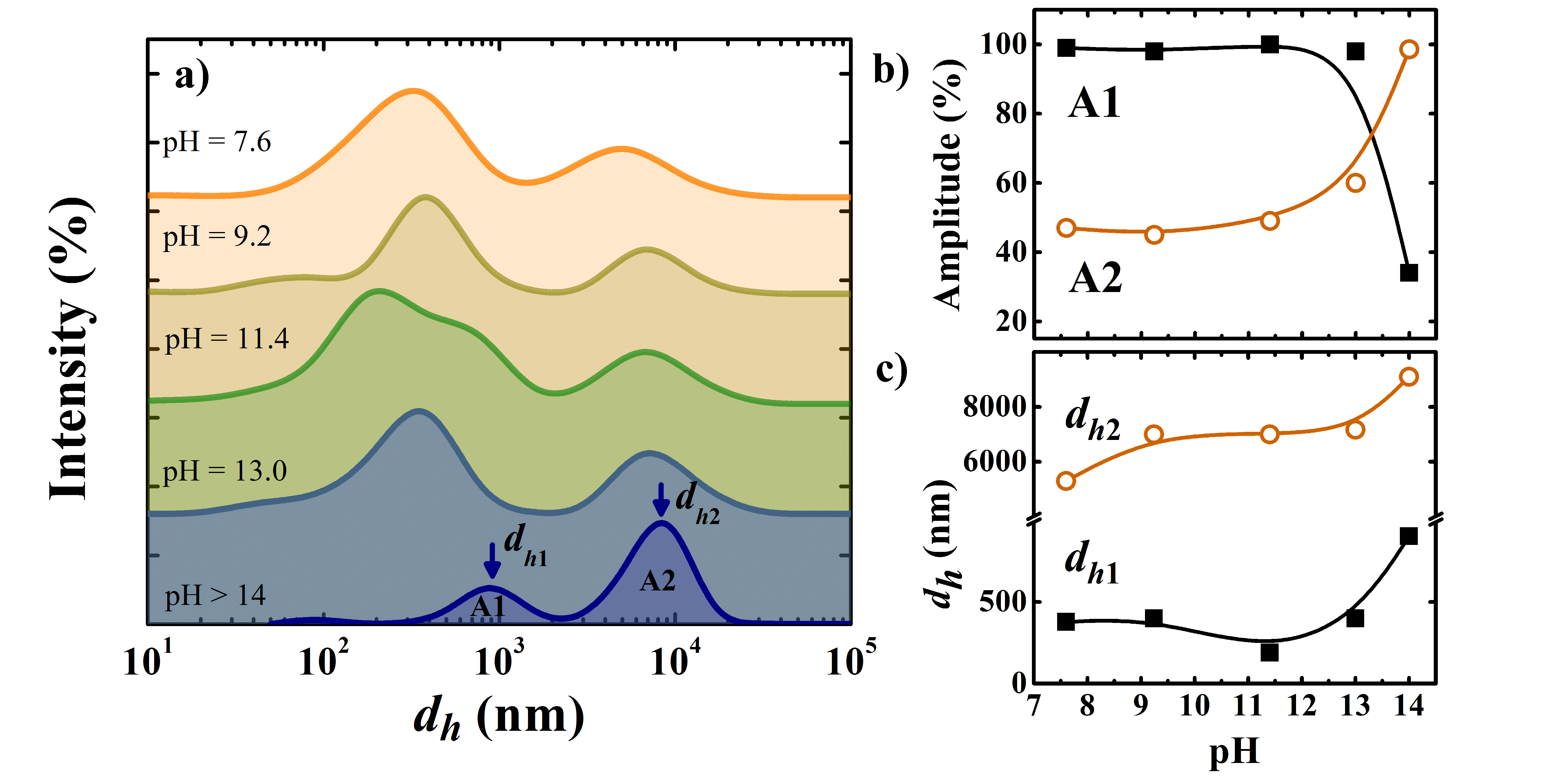}
	\caption{\textbf{(a)} Size distribution of the PANI beads (hydrodynamic diameter $d_h$) obtained using the CONTIN method to fit the autocorrelation functions obtained from the dynamic light scattering (DLS) measurements. \textbf{(b)} Amplitude and \textbf{(c)} hydrodynamic diameter as a function of the pH of the solution, showing that the effective size of the PANI particles can be modulated by this parameter.}
	\label{fig:Fig2}
\end{figure*}

The association of PANI with stabilizers, such as surfactants or other polymers, can produce supramolecular structures that protect the polymer against medium degradation without changing their conductivity characteristics. Plates, rods \cite{zhou2009synthesis}, tubes \cite{stejskal2022adsorption, zhou2008dilute}, gels \cite{wu2021reinforced}, and particles \cite{stejskal1996polyaniline, riede1998polyaniline} are examples of structures that can be constructed using PANI. Furthermore, low size dispersity polyaniline particles can be synthesized by a polymerization reaction of the monomer aniline, usually in acid medium, using many types of stabilizers such as polyvinylalchool (PVA), polyvinylpyrrolidone (PVP), colloidal silica \cite{stejskal1996polyaniline, riede1998polyaniline}, and sodium dodecyl sulfate (SDS) \cite{kim2001synthesis}.

In the present work we prepared spherical-shaped PANI particles using PVP as stabilizer with the goal to propose the use of this material in optical tweezers experiments. PVP was used since it has a proven efficiency compared to other surfactants \cite{stejskal1996polyaniline, riede1998polyaniline} probably due to the pyrrolidone group which has a favorable interaction with the polyaniline. We have shown that, depending on the synthesis parameters, micrometer-sized PANI particles can be trapped and manipulated using Gaussian beam OTs with a high stability. Furthermore, we characterized how the trap stiffness ($\kappa$) varies with some important parameters: the bead radius, the laser power and the bead height (distance from the bead center to the coverslip of the sample chamber) in pH 14. Tests performed under variable conditions have shown that this pH is the best one for the experiments described here, with the PANI beads presenting a semiconductive character that allows the stable trapping and, in addition, also optimum sizes in the micrometer range without significant aggregation (which occurs more frequently for lower pHs). For acidic pHs in fact the stable trapping cannot be achieved in our setup because the conductive character of the material is enhanced and the particles tend to form larger aggregates. To the best of our knowledge, there is not a previous study concerning the optical trapping and manipulation of organic semiconductor particles such as PANI. Therefore, the present work advances by explicitly demonstrating the viability in using this type of material in optical tweezers assays.

\begin{figure*}[ht]\centering 
	\includegraphics[width=\linewidth]{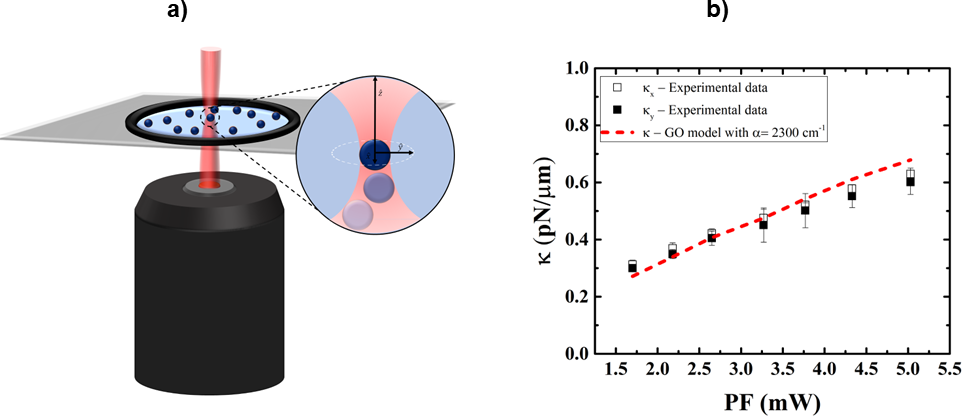}
	\caption{\textbf{(a)} Schematic illustration of the optical tweezers setup and sample chamber. \textbf{(b)} \textit{Squares:} transverse trap stiffness $\kappa$ for two perpendicular directions ($x$ and $y$) as a function of the laser power at the focus $PF$. \textit{Dashed red line}: aberration-free geometrical optics (GO) prediction for beads with an absorption coefficient $\alpha$ = 2300 cm$^{-1}$. Both experimental and theoretical data were obtained for a bead with radius  $a$ = (3.17 $\pm$ 0.02) $\mu$m at a height $h$ = 10 $\mu$m above the coverslip. Vertical error bars in the experimental data points are the calculated standard error of the mean obtained from a series of various measurements performed under the same conditions.}
	\label{fig:Fig3}
\end{figure*}

\section*{Materials and Methods}
\subsection*{Synthesis and characterization of the PANI microparticles}

Our PANI particles were synthesized based on procedures previously reported in the literature \cite{stejskal1996polyaniline, riede1998polyaniline}, according to the following steps: 2.5 mL of 1 M ammonium persulfate (APS) aqueous solution was slowly added to 10 mL of an aqueous solution with 0.072 mM of polyvinylpyrrolidone (PVP), 0.26 M of aniline and 97 mM of HCl. During the whole process the solutions were kept in an ambient with low luminosity and wrapped in aluminum paper. The temperature was maintained at 0 $^\circ$C. The solution was stirred with a magnetic bar during the addition of the initializer (1 hour) and left stirring for more 4 hours. After 2 hours a characteristic dark green color could be noticed. From this step, the sample remained at rest for 24 hours.

\begin{figure*}[ht]\centering 
	\includegraphics[width=\linewidth]{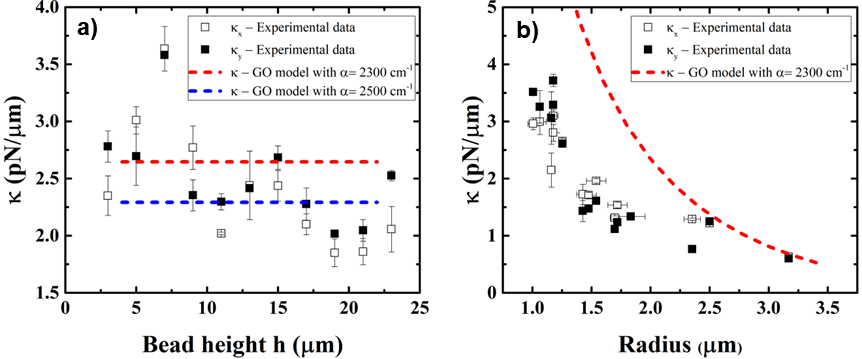}
	\caption{\textbf{(a)} \textit{Squares:} transverse trap stiffness $\kappa$ for two perpendicular directions ($x$ and $y$) as a function of the bead height $h$ relative to the cover slip. \textit{Dashed red line}: aberration-free geometrical optics (GO) prediction for beads with an absorption coefficient $\alpha$ = 2300 cm$^{-1}$. \textit{Dashed blue line}: similar theoretical prediction for $\alpha$ = 2500 cm$^{-1}$. Both experimental and theoretical data were obtained for a laser power at the focus of $PF$ = 9.65 mW and bead radius $a$ = (2.50 $\pm$ 0.03)  $\mu$m.  \textbf{(b)} Same type of data and theoretical calculations of panel (a), now obtained by varying the bead radius and maintaining a fixed height $h$ = 10 $\mu$m and a fixed laser power at the focus $PF$ = 4.97 mW. Vertical error bars in the experimental data points are the calculated standard error of the mean obtained from a series of various measurements performed under the same conditions.}
	\label{fig:Fig4}
\end{figure*}

For the optical tweezers samples, 200 $\mu$L of the previous solution was dispersed in a NaOH solution (2.5 mM) in a 1:5 (V/V) ratio, which has a pH $>$ 14. The resulting solution was stirred and sonicated for 8 minutes. The PANI particles were also examined by scanning electron microscopy (JEOL, model JSM-6010LA with an Energy Dispersive X-ray Spectroscopy detector with resolution equals to 133 eV) in order to confirm their spherical shapes and composition. In this case, the beads resulted from the synthesis process were separated from the solvent using a paper filter and washed with deionized water and acetone to remove unreacted aniline, PVP and APS before being imaged. A typical result obtained from this type of analysis is shown in Fig. \ref{fig:Fig1}. 

The particles composition was verified by Energy-Dispersion Spectroscopy (Fig. \ref{fig:Fig1} (c)) and the presence of carbon, nitrogen and oxygen were detected. Some sulfur were also present and are probably due to some APS that remained in the particle structure. Raman Spectroscopy measurements were performed (MicroRaman InVia Renishaw) using an argon ion laser ($\lambda$ = 514 nm) and microscope objective with 50x of magnification. The particles composition was again verified by Fig. \ref{fig:Fig1} (d), where the Raman spectra of PANI particles are compared to the one of PANI in emeraldine base form. The composition of the particles is again verified due to the presence of the peaks: 1598 cm$^{-1}$ (C=C stretching); 1510 cm$^{-1}$ (N-H deformation); 1328 cm$^{-1}$ (C–C stretching mode of the quinoid ring); 1206 cm$^{-1}$ (C-N stretching); 1170 cm$^{-1}$ (C-H deformation); 780 and 751 cm$^{-1}$ (ring deformation in the PANI emeraldine base); 533 and 417 cm$^{-1}$ (out-of-plane ring deformations in the PANI emeraldine base).

We have also performed dynamic light scattering (DLS) measurements in order to characterize the size distribution of the PANI particles. In these assays, the scattered intensity correlagrams were obtained from solutions composed of PANI particles re-dispersed in NaOH or HCl aqueous solutions at different pHs (7.6, 9.2, 11.4, 13.0, and $>$ 14), sonicated for 8 minutes. These measurements allowed us to investigate the size dependence of the particles with different degrees of doping achieved by changing the solution pH. The scattered intensity autocorrelation functions were fitted using the CONTIN method to obtain the intensity size distribution  shown in Fig. \ref{fig:Fig2}, panel (a). Independently of the pH used, it can be seen that there are typically two populations: a smaller one with hundredths of nanometer (population 1) and a larger one with a diameter one to two orders of magnitude higher (population 2). The large particles have diameters of micrometers which seems unlikely a reliable result using DLS and could be interpreted as an artifact come from dust or impurity, for instance. Nonetheless the intensity autocorrelation function of all samples shown a smooth aspect and gave the same results after some repetitions (see Supporting Information). Even if the large diameters are not reliable by these experiments (the diffusivity of the particles are probably overestimated by process of sinking) the diameters order of magnitude and dependence with changing the pH can be trusted. When the pH of the solution decreases, the intensity percentage of population 2 decreases. Furthermore, the hydrodynamic diameter of both populations tend to decrease by reducing the pH (Fig. \ref{fig:Fig2}, panel (c)). This effect could be explained by the protonation of the PANI chains, which is lower or nonexistent at higher pHs but significant at lower pHs, increasing the chains solubility. In Fig. \ref{fig:Fig2} (b) and (c) it is shown the amplitude (A1 and A2) and the peak diameter (d1 and d2) of the two populations.

\subsection*{Optical Tweezers setup}

The optical tweezers (OT) consist of a linearly polarized infrared laser operating in the usual Gaussian ($TEM_{00}$) mode, wavelength $\lambda$ =  1,064 nm, with a (8.80 $\pm$ 0.08) mm beam waist measured at the objective entrance (whose aperture diameter is 6.0 mm). The tweezers is mounted on a Nikon Ti-S inverted microscope with a 100$\times$ N.A. 1.4 oil-immersion objective. The sample chamber is constructed by sandwiching an o-ring between two glass coverslips, and the colloidal PANI solution is added inside this o-ring (see Fig. \ref{fig:Fig3}(a)).

\section*{Results and Discussions}

In all OT assays the specific hydrodynamic radius of each bead chosen was measured via the diffusion coefficient of the particle when free in solution, determined by videomicroscopy. The trap stiffness $\kappa$ was measured by analyzing the Brownian fluctuations of the bead position in the optical potential for one minute, using a CMOS camera (Basler ac1920-155uc). The technical details of our analysis process were described elsewhere \cite{andrade2021bessel}.

In Fig. \ref{fig:Fig3}, panel (b), we show the experimental trap stiffness (\textit{squares}) measured as a function of the laser power at the focus (\textit{PF}) for the two axes ($x$ and $y$) transverse to the laser propagation direction ($z$). Such measurements were performed using a PANI bead with a radius $a$ = (3.17 $\pm$ 0.02) $\mu m$ and at a height of 10 $\mu m$ above the surface of the bottom coverslip of the sample chamber.

The \textit{dashed red line} shown is the theoretical result predicted by a geometrical optics (GO) model previously developed in our group \cite{aberracaorocha2009optical}. Such model is based on the Ashkin's original model for the radiative force on spherical particles \cite{GOashkin1992forces}, generalized to include the light absorption by the particles via the absorption coefficient of the material \cite{absortioncampos2018light}. In the present case, the version of the model that neglects spherical aberration provided a better agreement with our experimental data, and all GO calculations presented in this manuscript were thus done under this situation. The reason is related to the non-negligible absorption of the particles, which limits GO calculations especially when computing optical aberrations, as will be evident later. Complete details of the GO calculations can be found in the original references \cite{aberracaorocha2009optical, absortioncampos2018light}. It is worth to say that the comparison between theory and experiments shown in Fig. \ref{fig:Fig3} (as well as those shown in the next figures) is absolute since the theoretical data was obtained without fitting parameters. In fact, most of the parameters needed to calculate the theoretical trap stiffness were carefully measured; they are the bead radius and height from the coverslip and the laser power and beam waist (the refractive index of PANI (1.8) was found in the literature \cite{Farag}). In addition, for PANI we have used an absorption coefficient  $\alpha$ = 2300 cm$^{-1}$, compatible with the literature \cite{de2021photoconductivity}. Such value of $\alpha$ provides the best agreement between the GO model and our experimental results, indicating that the PANI beads absorb considerably at the 1,064 nm wavelength used. 

Absorbent beads in optical traps can exhibit different behaviors. When the absorption coefficient is compatible with those of metals ($\alpha_{metals} \simeq$ 10$^{5}$ cm$^{-1}$ \cite{johnson1972optical}), the absorption of light usually generates a photophoretic force that dominates the gradient force largely, preventing micrometer-sized beads from being trapped. When the absorption coefficient is much smaller than that of metals, on the other hand, the effect of light absorption is to weaken the trap stiffness \cite{absortioncampos2018light}.

In Fig. \ref{fig:Fig3}, note that the experimental data of the trap stiffness slightly deviates from the classic linear behavior as a function of the laser power. Furthermore, for the lowest powers used the experimental data is slight higher than the theoretical prediction, and the opposite behavior occurs for the highest powers used. Such type of behavior was previously verified for other absorbing beads (although for higher laser powers) \cite{andrade2021bessel} and strongly suggests that the absorption coefficient of PANI beads is intensity-dependent, as in the case of traditional semiconductors such as germanium or silicon \cite{meyer1980optical,meyer1980optical2}.
In other words, increasing laser intensity increases the rate of generation of free charge carriers in semiconductors, which in turn increases the absorption coefficient by the free carrier (FC) absorption mechanism ($\alpha_{FC}$) \cite{meyer1980optical,fox2002optical}. Thus, the deviation of the experimental data  from the linearity can be understood by assuming that for powers less than $\sim$ 3.5 mW the absorption coefficient is slightly lower than 2300 $cm^{-1}$ (and therefore the experimental $\kappa$ is higher than that predicted by the GO model); while for laser powers higher than 3.5 mW the absorption coefficient is slightly higher than 2300 $cm^{-1}$ (and therefore the experimental $\kappa$ is smaller than predicted by the GO model).

In Fig. \ref{fig:Fig4}, panel (a), we show the measured transverse ($x$ and $y$) trap stiffness $\kappa$ as a function of bead height $h$ (\textit{squares}). Such measurements were performed using a PANI bead with radius $a$ = (2.50 $\pm$ 0.02) $\mu$m and for a fixed laser power at the focus $PF$ = 9.65 mW. We also show the prediction of the GO model using the same absorption coefficient used in the Fig. \ref{fig:Fig3}(b), $\alpha$ = 2300 cm$^{-1}$ (\textit{dashed red line}) and also $\alpha$ = 2500 cm$^{-1}$ (\textit{dashed blue line}) for comparison purposes. We have included these two different $\alpha$ values to emphasize the effect of absorption dependence on the laser power: note that now the power at focus in this case is twice that used in Figure \ref{fig:Fig3}(b), such that it is expected that in this case the absorption by PANI particles is higher. In addition, one should note that the experimental data of $\kappa$ exhibits a slight decrease with increasing bead heights. The aberration-free GO model, on the other hand, predicts that $\kappa$ is constant as the bead height (\textit{h}) varies, as shown by the horizontal dashed lines in the Fig. \ref{fig:Fig4}(a). Spherical aberration in general degrades the laser focus, decreasing the trapping efficiency of optical tweezers constructed using Gaussian beams \cite{aberracaorocha2009optical,aberracaooliveira2018optical}. This result thus suggests that our system is not completely free from spherical aberration effects, although its effect is minimal.

Finally, we also investigated the behavior of $\kappa$ by varying the radius of the beads as shown Fig. \ref{fig:Fig4}(b). The GO model has already proved to be efficient in predicting the behavior of $\kappa$ as a function of the radius for dielectric polystyrene  beads in the size range $a > >\lambda$ up to $a\simeq \lambda$ \cite{aberracaorocha2009optical}. For PANI, our experimental data have the expected behavior for the variation of $\kappa$ as a function of the radius of the bead, an hyperbolic curve, but the values obtained are smaller than those predicted by the GO model for $a <$ 2.5 $\mu$m. This result suggests that, for absorbing beads, the GO model does not provide a good approximation for the trap stiffness when $a \simeq \lambda$, and more sophisticated models are needed to predict the behavior of $\kappa$, such as the MDSA+ model - which includes other optical aberrations \cite{Dutra2}; or, alternatively, the GLMT approach \cite{gouesbet1985scattering}.

The success of the GO model in predicting the variation of $\kappa$ as a function of the radius for transparent polystyrene beads when $a \simeq \lambda$ comes from the fact that the incident rays that contribute to the trap when the particle is located on the optical axis are not affected by dominant diffraction and resonance effects, well described only by Mie scattering \cite{viana2007towards}.  For absorbing beads, such as PANI, the type of scattering is completely different from the that found in transparent particles \cite{hulst1981light}, including, among other effects, the annihilation of the photons that are absorbed. Therefore, the Mie scattering effects become more relevant and the GO model starts to fail for $a \simeq \lambda$, in agreement with our experimental data (Fig. \ref{fig:Fig4}(b)). In any case, independently of a better model that can be used to predict the behavior of the trap stiffness as a function of the bead radius for the PANI particles, such quantity can be determined experimentally with high accuracy, which guarantees that PANI can be stably trapped in a Gaussian beam OT within a wide range of particle sizes. The results of the present work thus open the door for using new semicondutor polymeric materials as handles for optical tweezers applications.

Furthermore, observe that in the present situation (pH = 14 and low laser power used) the photophoretic effects are not relevant to the system. This conclusion can be achieved from the data of Fig. 4b. In fact, if photophoretic forces play a role in such situation, we would expect that the experimental data deviates from the GO model for higher particles radius, since photophoresis is more relevant for larger particles and the model does not compute such an effect. Nevertheless, our data of Fig 4b present the opposite behavior, with a convergence between the experimental and the theoretical results for larger particles. Such a result strongly indicates that photophoretic forces are not relevant to the present system at the experimental conditions used, although this fact can changes drastically under other situations (lower pHs, higher laser powers, etc).

\section*{Conclusion}

In summary, we have successfully synthesized spherical PANI particles with a wide size range (1 - 3.5 $\mu$m radius), ideal for optical tweezers experiments. A robust characterization of the optical trapping of such particles was performed in a Gaussian (TEM$_{00}$) beam optical tweezers, attesting the viability in using this type of semiconductor polymeric material as handles for this technique. The results found show that the PANI particles behave like a inorganic semiconductor with an absorption coefficient that tends to increase with the laser power. Furthermore, it is also expected that such coefficient depends on the pH solution \cite{paniph}, and on the rate of PANI/stabilizer used to produce the beads \cite{mansour2015synthesis}. These characteristics suggest the optical manipulation of PANI particles can be strongly modulated, and the possibility to obtain an oscillatory behavior similar to that observed with inorganic semiconductors \cite{campos2019germanium,moura2020silicon} cannot be discarded. In addition, due to the fact that those particles are made of an organic semiconductor, they can be functionalized with other polymers and/or molecules more easily than inorganic semiconductors for specific applications in microrheology and other fields. Finally, as a relevant perspective of this work it is worth to say that, as a consequence of the semiconductor behavior of the PANI particles, the optical forces on them can be modulated \cite{moura2020silicon} through the charge carriers generation, optimizing for example the construction of photophoretic traps. Moreover, the possibility of modulating the absorption coefficient allows the building of these types of traps at higher wavelengths (around infrared), opening the door for improvements in technological applications, \textit{e. g.}, to increase the efficiency of volumetric displays, whose construction is based on photophoretic traps \cite{shvedov2011robust}.


\section*{Supporting Information}
The Supporting Information is available free of charge at link.
The supporting information have more details about the DLS measurements and a video showing a typical optical tweezers experiment.


\phantomsection
\section*{Acknowledgments} 


This research was funded by Conselho Nacional de Desenvolvimento Cient\'ifico e Tecnol\'ogico (CNPq); Financiadora de Estudos e Projetos (FINEP); Funda\c{c}\~ao de Amparo \`a Pesquisa do Estado de Minas Gerais (FAPEMIG); and Coordena\c{c}\~ao de Aperfei\c{c}oamento de Pessoal de N\'ivel Superior (CAPES) - Finance Code 001; and INCT of Spintronics and Advanced Magnetic Nanostructures (INCT-SpinNanoMag).

\section*{Conflict of Interest}
The authors declare no conflict of interest.

\section*{Data Availability Statement}
The data that support the findings of this study are available from the corresponding author upon reasonable request.


\phantomsection
\bibliographystyle{acm}

\bibliography{Article_PANI-OT}

\newpage


\newpage

\section*{Table of Contents Graphic (TOC graphic)}
\includegraphics[width=18cm]{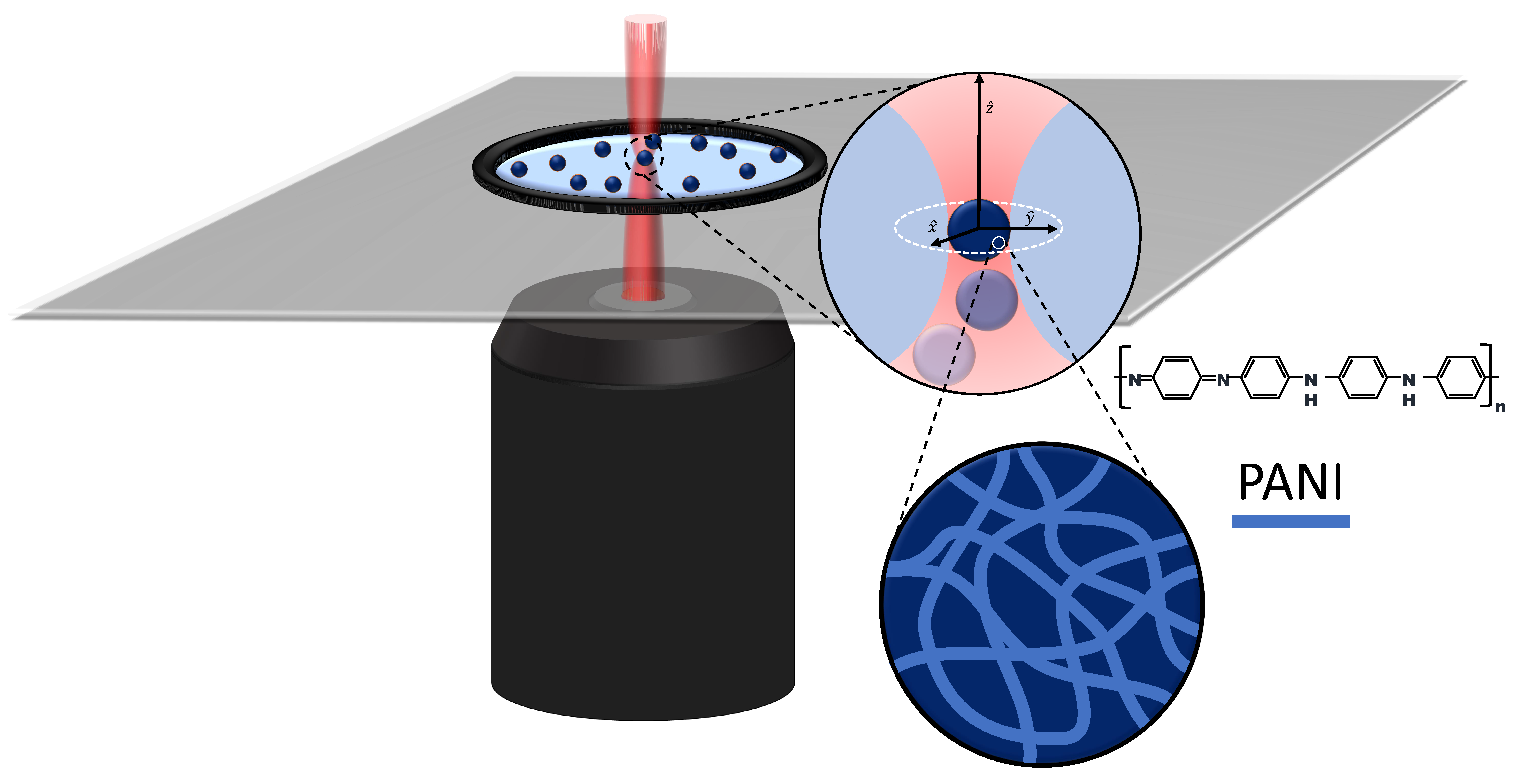}

\end{document}